# Improved chemical vapor transport growth of transition metal dichalcogenides

Alberto Ubaldini and Enrico Giannini

*Département de Physique de la Matière Condensée, University of Geneva, Switzerland*

ABSTRACT

In the crystal growth of transition metal dichalcogenides by the Chemical Vapor Transport method (CVT), the choice of the transport agent plays a key role. We have investigated the effect of various chemical elements and compounds on the growth of $TiSe_2$, $MoSe_2$, $TaS_2$ and $TaSe_2$ and found that pure $I_2$ is the most suitable for growing $TiSe_2$, whereas transition metal chlorides perform best with Mo- and Ta- chalcogenides. The use of $TaCl_5$ as a transport agent in the CVT process allows to selectively growth either polymorph of $TaS_2$ and $TaSe_2$ and the optimum growth conditions are reported. Moreover, by using $TaCl_5$ and tuning the temperature and the halogen starting ratio, it was possible to grow whiskers of the compounds $TaS_2$, $TaSe_2$, $TaTe_2$, $TaS_3$ and $TaSe_3$.

INTRODUCTION

Transition metals dichalcogenides $MX_2$ (where M is a transition metal and X a chalcogen element: O, S, Se and Te) are a very wide group of materials with a large spectrum of physical properties that make them very attractive.
Many of them, namely when the transition metal belongs to groups IVA, VA and VIA, or M = Rh, Ir, Pd, Pt, and X≠O, crystallize in quasi-2D layered structures, in which $MX_2$ slabs are weakly bonded by Van der Waals forces [1]. For this reason, exfoliation down to extremely thin flakes, even to single layers, is possible. Inside the layers the bonds have strong covalent and ionic nature.
This peculiar structure is responsible for electronic and mechanical properties that make these materials suitable for applications as superlubricants [2], ultralow thermal conductivity devices [3], catalyzers [4], and solar cell converters [5]. Very recently, it has been demonstrated that some semiconducting $MX_2$ with an excellent electronic mobility, are suitable for the fabrication of electronic devices, such as transistors, [6,7]. For these reasons it is extremely important to control the growth of high quality single crystals of these materials. In most cases these materials are grown by the chemical vapor transport method (CVT). The basic principles of this method are long known: a thermal gradient drives the crystallization from a vapor, formed from a source of raw materials heated up to high temperature. If the vapor partial pressure of one or more components is too low, (for practical purposes, a relevant transport occurs only when $p \geq 0.1 * 10^{-3}$ mbar [8]) a chemical substance is added to the initial mixture in order to favor the formation of more volatile species. For this process, the equilibrium constants of the reaction between the non-volatile material and the transport agent should have an appropriate value. If it is too low, no reaction occurs and no transport is possible. On the contrary, if it is too high, the new species formed are too stable, they will not decompose and the growth would be inhibited.
The choice of the correct vapor transport agent is therefore one of the critical points of this method. The other key parameters that must be controlled are the temperatures of the two extremities and the molar ratio between the metal and the vapor transport agent. Out of the narrow range of temperature and composition that favors the growth of bulk crystals, it is possible to grow low-dimensional structures, particularly one-dimensional whiskers, of these compounds.
In this work, we report a study on the chemical nature of the transport agents for $TiSe_2$, an interesting material that can exhibit coexistence of superconductivity and charge density waves (CDW) [9], and $MoSe_2$, one of the most promising semiconducting dichalcogenides [7]. For the former, the best results are obtained by using iodine as a transport agent, whereas for the latter by using $MoCl_5$. A similar reagent, $TaCl_5$, is found to be effective for growing $TaX_2$ crystals. In this case, each possible polymorphic phase (either the trigonal "2H" with prismatic coordination or the orthorhombic "1T" with octahedral coordination) can be selectively grown by carefully controlling the growth conditions.
Finally, we also present the growth of $TaX_2$ and $TaX_3$ whiskers, grown by the vapor transport method based on the use of metal chloride as source.

EXPERIMENTAL

Single crystals of $TiSe_2$ were grown using four different transport agents: $I_2$, $NH_4Cl$, $NH_4Br$ and $NH_4I$. The transport agent was added to a stoichiometric mixture of titanium and selenium and sealed under vacuum ($10^{-5}$ torr) inside a quartz tube. The molar ratio between $I_2$ and Ti was set to 0.05, whereas for the others the ratio between the halogen and Ti was twice as much, in order to have the same molar fraction of the halogen elements. The amount of titanium, as well all the parameters (T, ΔT, heating rate, and reactor volume), were the same in all cases. These mixtures were heated for 24 h at $T_{hot}$ = 680°C, whereas the end of the tube that does not contain the precursors was maintained at lower temperature ($T_{cold}$ = 600°C). The length of the quartz reactor being ~10-12 cm, a temperature gradient of ~6-8°C/cm was maintained between the hot and the cold ends.

A similar procedure was used for $MoSe_2$, using a ratio between $I_2$ and Mo as high as 0.1, and 0.2 for the other transport agents. Moreover, $MoCl_5$ was also tested as possible source of chlorine. In this last case, the mixture was prepared according to the reaction:

$$14/15\ Mo + 1/15\ MoCl_5 + 2\ Se \rightarrow MoSe_2 + 1/6\ Cl_2. \qquad\qquad 1)$$

The reaction temperature of the precursors was 820°C, and the cold end of the tube was kept at 730°C, for a growth time of 24 hours.

Crystals of $TaX_2$ were grown as well, using $TaCl_5$ as a source, according to a similar reaction, but with a different initial fraction of $TaCl_5$. The $TaCl_5$:Ta ratio was tuned depending on the chalcogen element X and on the wished crystal structure. For stabilizing the 2H structure the highest temperature was 720°C, 820°C and 850°C for X = S, Se and Te, respectively ($T_{cold}$ = 650°C, 730°C, and 750°C, respectively). The temperature was set to 1050 °C ($T_{cold}$ = 920°C) for the growth of the 1T structure.

In addition to layered bulk crystals, we succeeded in growing one dimensional whiskers of $TaX_2$ as well. For this aim the growth temperature was lowered and the initial quantity of $TaCl_5$ was increased. In particular, these structures grew at $T_{cold}$ = 550, 580 and 600°C for for X = S, Se and Te, respectively. This method resulted also to be effective for the growth of $TaS_3$ and $TaSe_3$ whiskers, prepared from a starting ratio Ta:X = 1:3 at low temperature, $T_{cold}$ = 500 and 520°C, respectively.

The single crystals were mechanically extracted from a polycrystalline agglomerate with a typical desert-rose shape. The samples were first characterized by X-ray diffraction (XRD), in a four-circle diffractometer Siemens D5000 using the CuKα radiation and in a Multiwire Digital Laue camera. SEM-EDX analyses were performed both in a LEO 438VP coupled to a Noran Pioneer X-ray detector, and in the high-resolution microscope JEOL JSM 7600F.

RESULTS AND DISCUSSION

Figure 1 (a-d) shows various $TiSe_2$ crystals grown using different transport agents under the same growth conditions, $T_{hot}$ = 680°C, $T_{cold}$ = 600°C, Δt = 24 hours. Within such a short reaction time, the effect of the chemical nature of the transport agent at the early stage of growth is highlighted. Isolated bulk crystals were found to form only in the presence of $I_2$ (Fig. 1(d)), whereas crystalline aggregates formed by using $NH_4I$ (Fig. 1(c)). The other salts of ammonium led rather to powders instead of crystals, the smallest grains being observed when using ammonium chloride.

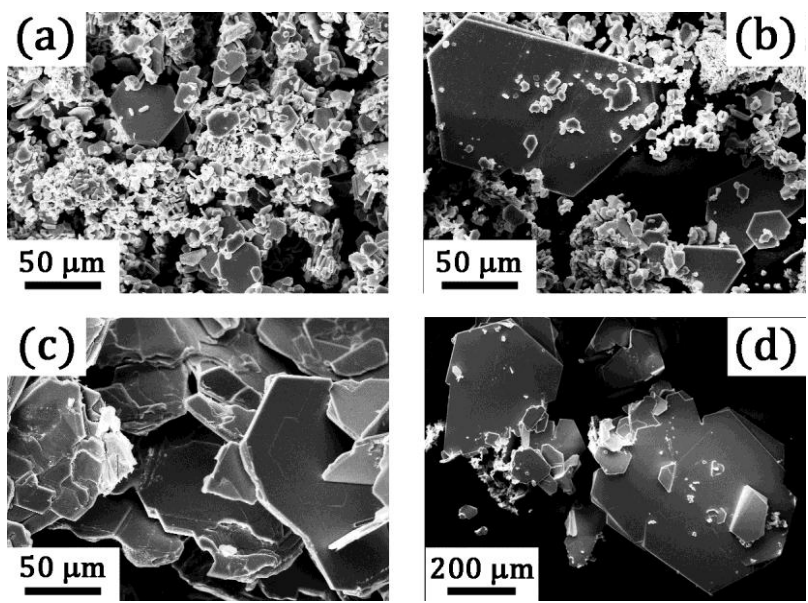

*Fig. 1: crystals of $TiSe_2$ grown in 24 h at $T_{hot}$ = 680 °C, using $NH_4Cl$ (a), $NH_4Br$ (b), $NH_4I$ (c) and $I_2$ (d) as vapor transport agent.*

Large single crystals of $TiSe_2$ grown using $I_2$ as transport agent are shown in Fig. 2. The upper left panel, Fig. 2(a), shows a picture of 3-4 mm wide crystals with shiny surfaces and sharp 120° angles. The upper right panel, Fig. 2(b), shows a Laue pattern of one of these crystals, aligned along the [*001*] plane (the cleavage plane). In the lower panel, Fig. 2(c), the θ-2θ diffraction pattern shows only narrow [*00l*] reflections and the inset shows a [*002*] rocking curve with a FWHM as narrow as ~0.19°. These results from XRD prove the high crystalline quality of these samples. EDX analyses (not shown for the sake of conciseness) have confirmed a homogeneous $TiSe_2$ composition.

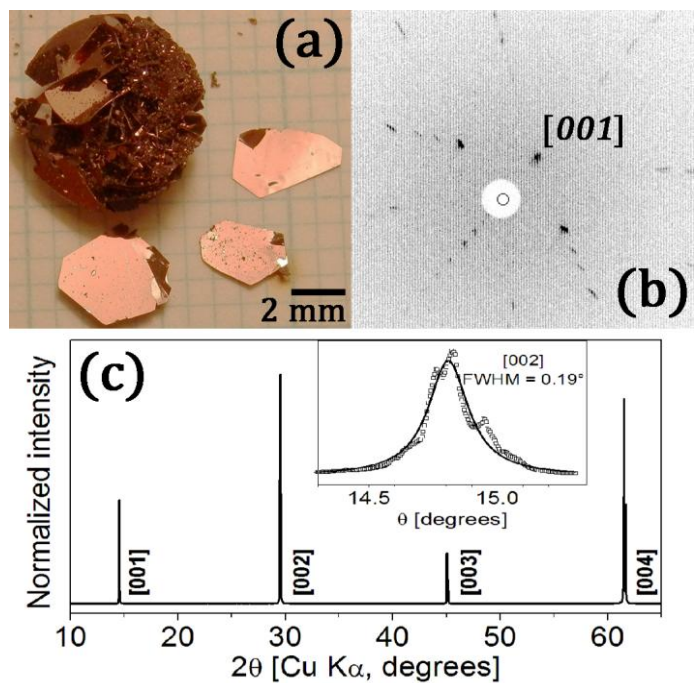

*Fig. 2: (a) optical picture of large crystals of $TiSe_2$; (b) Laue diffraction pattern of a single crystal of $TiSe_2$ oriented along [001]; (c) X-ray diffraction pattern on a $TiSe_2$ crystal showing only narrow [00l] reflections with a typical [002] rocking curve (FWHM=0.19°) shown in the inset of panel (c).*

The growth process passes through an initial reaction between titanium and the halogen that produces many volatile halides, above all $TiX_4$ [10, 11]. For the formation of these molecular species from the ammonium salts, the preliminary step of thermal decomposition is necessary: $NH_4Y \rightarrow NH_3 + HY$ (Y = halogen element). The acid formed in this way can in turn react with titanium ($Ti + 4\ HY \rightarrow TiY_4 + 2\ H_2$) and the metal transport can occur, thus feeding the growth process. However, the concentration of HY, and therefore the reaction rate, is expected to be low at the beginning of the process. Since the decomposition temperature decreases as the atomic number of the halogen increases, the partial pressure of the volatile chlorides is lower than that of the other halogenides, thus making $NH_4Cl$ the least effective transport agent. For the same reason, $NH_4Br$ is expected to be less effective than $NH_4I$. The higher effectiveness of pure iodine, compared to ammonium iodide, relies on the low temperature at which the direct sublimation reaction of Ti occurs. As a matter of fact, the thermal decomposition of ammonium halides creates other molecular species that, despite not directly involved in the reactions, change the pressure, thus making the partial pressure of the volatile species lower. Unintentional doping due to the presence of halogen atoms was found not to occur within the accuracy of our EDX analysis.

For $MoSe_2$, the results are different, at least using the present growth conditions: large crystals are found only using $MoCl_5$ as a source, whereas by using the ammonium salts practically no growth occurs and only few and very small crystals form when using iodine. Increasing the starting amount of the transport agent would make it difficult to control the nucleation rate, and therefore would not have a beneficial effect on crystal growth. The reason for such different behavior between Ti and Mo, should be looked for in the different thermodynamic equilibriums establishing between the metal and the transport agent. Our results show that Ti reacts more easily with iodine and with the other halogens, than molybdenum. As a consequence, the vapor pressure of the Mo-carrying volatile compounds remains very low and the process cannot proceed effectively. On the contrary, $MoCl_5$ sublimates at low temperature [12] and provides a high enough partial pressure of Mo-containing species in the vapor, which in turn enhances the growth rate.

The selection of the vapor transport agent is therefore a critical point that must be carefully taken into account for growing crystals of transition metal dichalcogenides. We have recently investigated the use of transition metal chlorides as effective transport agents for growing some of these compounds [13].

Similarly to $MoSe_2$, $TaX_2$ crystals were grown using $TaCl_5$. Differently from the aforementioned $TiSe_2$ and $MoSe_2$, two stable polymorphs of TaS$_2$ and TaSe$_2$, the so-called "1T" and "2H" structures, exist in distinct temperature ranges. The 2H structure (P6$_3$/mmc space group, NbS$_2$ structure type) is thermodynamically stable at low temperature (T < 840°C in the Ta-Se system) and the 1T structure (P-3m1 space group, CdI$_2$ structure type) at high temperature [14]. In this work, we show that, thanks to the improved CVT method based on chlorides, it is possible to grow selectively either polymorph of this compound, by simply tuning the temperature of the growth zone, either higher or lower than the phase transition temperature.

Figure 3(a–c) shows crystals of 2H (Fig. 3(a)) and 1T (Fig. 3(b-c)) TaS$_2$, grown at T$_{hot}$ = 720°C and 1050°C, respectively (T$_{cold}$ = 650°C and 920°C, respectively), using different amounts of chloride. Figure 3(d) shows the XRD patterns of crystals corresponding to the samples in figures 3(a) and 3(c) and confirms the different crystal structures. The chemical composition of these crystals was found to be very homogeneous by EDX analyses.

The initial TaCl$_5$ concentration is a key parameter for controlling the transport, nucleation and growth rates. A high chloride fraction leads to the formation of an extremely high number of very small crystals and eventually to a thick polycrystalline layer, whereas a too low one does not create the partial pressure needed for an effective transport process. It was found that the optimal ratio between TaCl$_5$ and Ta is between 1/50 and 1/30 [13]. The crystals shown in figure 3(a) were grown using a ratio of 1/50.

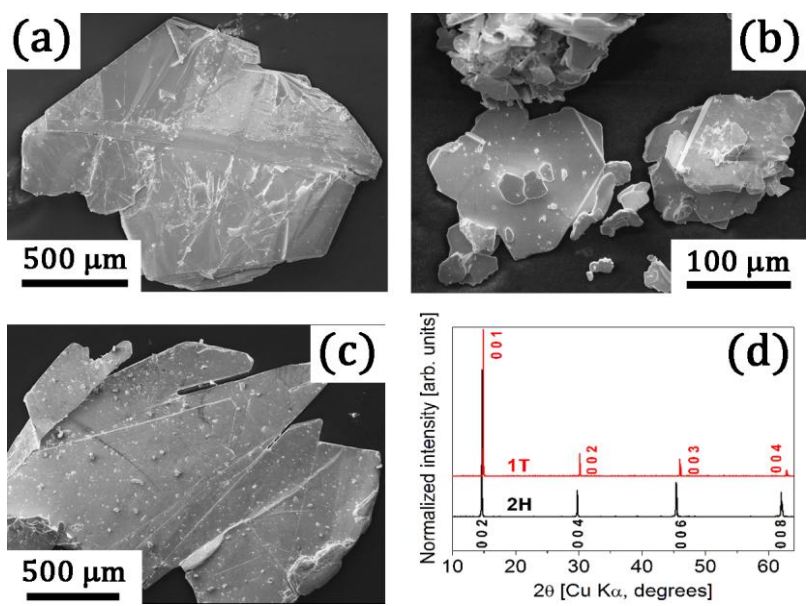

*Fig. 3: crystals of 2H – TaS$_2$ (**a**) prepared at T$_{hot}$ = 720 °C, T$_{cold}$ = 650°C, in 24 h, and of 1T – TaS$_2$ prepared at T$_{hot}$ = 1050 °C, T$_{cold}$ = 920°C, in 24 h using TaCl$_5$:Ta = 1:80 (**b**) and TaCl$_5$:Ta = 1:30 (**c**). Panel (**c**) shows XRD patterns of samples a and c.*

The growth conditions found for the 2H phase would suggest that the TaCl$_5$:Ta ratio should be further lowered in order to grow the 1T phase that forms at higher temperatures, thus keeping the transport and growth rates slow enough. Surprisingly, it was found that lowering the initial chloride fraction leads to the formation of smaller crystals (see Fig. 3(b)). Moreover, and more surprisingly, the crystals of the 1T phase did not grow in the cold end of the quartz ampoule but form an aggregate on the hot side. According to the mechanism above explained, this is counter-intuitive and a different growth process has to be invoked. A possible interpretation requires that either the nucleation or the growth process is very slow. If the transport in the vapor phase is very effective and fast, because of the high temperature, then the homogeneous nucleation of the 1T phase at the cold end must be prevented and the nuclei appear preferentially on the surface of the metallic pieces of Ta in the hot part of the reactor. Once they are formed, the crystal growth is fed by a fast matter supply. In an alternative scenario, it is possible that at very high temperature the gaseous chlorides begin to dissociate, as it has been observed for other halogenides [15], and the actual partial pressure is very low. In such situation the transport would not be effective, and crystals would grow close to the precursor source at a rather high rate.

The last possibility offered by the use of chlorides as a source is the growth of 1D-structures. Figure 4(a-c) shows whiskers of TaS$_2$, TaSe$_2$ and TaTe$_2$. To the best of our knowledge, whiskers of TaTe$_2$ have not been reported previously. These quasi 1D structures can grow thanks to a vapor – solid (VS) mechanism as described elsewhere [13], and reported previously for other systems [16–19]. Very often low supersaturation is more suitable for the growth of 1-D structures than for bulk structures. Such condition is satisfied when the growth temperature is particularly low.

These whiskers have a large aspect ratio, being few mm long and only hundreds nm wide. They are rather flat then cylindrical, and the largest exhibit several terraces on the surface. In order to achieve an effective matter transport at such low temperature, the initial amount of TaCl$_5$ had to be increased up to 1/5. Even under these conditions it is possible that layered bulk crystals still grow together with whiskers. Shorter and less numerous whiskers are obtained for TaTe$_2$ (Fig. 4(d)), suggesting that the growth conditions are not optimized yet. In the systems Ta – S and Ta – Se, the phases TaS$_3$ and TaSe$_3$ also exist and are stable at lower temperature than the dichalcogenides. Using the chloride driven method it is possible to grow whiskers of these compounds as well (Fig. 4(d)). They are obtained starting from an initial mixture enriched in sulfur and selenium, according to the stoichiometric ratio, and at even lower temperatures.

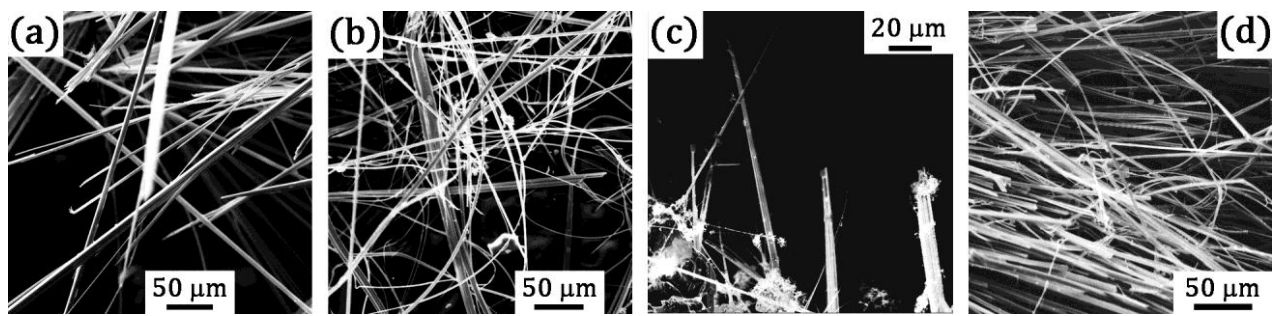

Fig. 4: whiskers of TaS$_2$ **(a)**, TaSe$_2$ **(b)**, TaTe$_2$ **(c)** and TaS$_3$ **(d)** grown using TaCl$_5$ as source.

CONCLUSIONS

The chemical vapor transport method is effective for crystal growth of many layered chalcogenide materials. The correct choice of the transport agent depends on the particular compound to be grown and optimizes the growth kinetics and the final crystal quality. For the TiSe$_2$ good results are found using iodine, but other transport agents are needed in the presence of Mo and Ta. In these cases, large crystals are grown using MoCl$_5$ and TaCl$_5$ as precursors. Tantalum disulfide and diselenide can crystallize in various polymorphic structures that can be selectively grown by controlling the growth temperature. The same method has made it possible to grow whiskers of these materials.